\documentstyle[12pt,prl,aps,psfig]{revtex}
\newcommand{\be}{\begin{equation}}
\newcommand{\ee}{\end{equation}}
\newcommand{\la}{\langle}
\newcommand{\ra}{\rangle}
\newcommand{\bea}{\begin{eqnarray}}
\newcommand{\eea}{\end{eqnarray}}
\newcommand{\om}{\omega}

\newcommand{\br}{{\bf r}}

\newcommand{\eps}{\epsilon}
\newcommand{\pa}{\partial}
\begin{document}
\draft
\title{Effects of electrostatic fields and Casimir force\\ on cantilever
vibrations}
\author{A. A. Chumak$^{1,2}$,  P. W. Milonni$^1$, and G. P. Berman$^1$}
\address{$^1$ Theoretical Division, Los Alamos National Laboratory, Los Alamos,
New Mexico 87545}
\address{$^2$ National Academy of Sciences of Ukraine, Institute of Physics,
\\Pr. Nauki, 46, Kiev 03650, Ukraine} 
\maketitle
\begin{abstract}
    The effect of an external bias voltage and fluctuating electromagnetic
fields on both the fundamental frequency and damping of cantilever vibrations
is considered. 
An external voltage induces surface charges causing cantilever-sample
electrostatic attraction. 
A similar effect arises from charged defects in dielectrics that cause spatial
fluctuations of electrostatic fields. The cantilever motion results in charge
displacements giving rise to Joule losses and
 damping. It is shown that the dissipation increases with decreasing
conductivity and thickness of the substrate, a result that is potentially 
useful for sample diagnostics. 
Fluctuating electromagnetic fields between the two surfaces also induce
attractive (Casimir) forces. It is shown that the shift in the cantilever
fundamental frequency due to the
Casimir force is close to the shift observed in recent experiments of Stipe {\it
et al.} \cite{ru}. Both the electrostatic and Casimir forces have a strong
effect on the 
cantilever eigenfrequencies, and both effects 
depend on the geometry of the cantilever tip. We consider cylindrical,
spherical, and ellipsoidal tips moving parallel to a flat sample surface. The
dependence of the cantilever 
effective mass and vibrational frequencies on the geometry of the tip is studied
both numerically and analytically. 
\end{abstract}
\section{Introduction}
The description of the motion of a microscopically small body near a massive
solid is an important problem in various areas of modern physics, including
atomic, scanning tunneling, and magnetic resonance force microscopies. The
interactions between bodies separated by a vacuum arise from electromagnetic
fields 
of various types, the simplest example being an external potential difference
applied between metallic bodies. In this case the surfaces in close proximity
acquire opposite electric charges 
and experience an electrostatic attraction. A static electric field between
different surfaces may exist even without any externally applied voltage; a
zero-bias electrostatic attraction may 
be caused by local variations in the work function or surface contamination
resulting in an inhomogeneous electric field. This is referred to as the patch
effect (see, for instance,
References \cite{young}-\cite{speake}). Also, spatial fluctuations of the
electric field may be caused by charged defects in the bulk of a cantilever or
a sample. In addition to these electrostatic effects, electrically neutral
bodies interact via fluctuating electromagnetic fields. Fluctuating
 electric currents and charge densities (or polarizations) of a given body
produce fields that act on other bodies. The strength of the interaction
depends on the power spectrum of 
the fluctuations, which in turn are determined by the complex dielectric
permittivity. Lifshitz treated this van der Waals force in a classic paper
\cite{lif}. 
In the case of high-conductivity metal surfaces the effect is referred to as the
Casimir force, which has been the subject of several 
recent experiments \cite{lamor}-\cite{onofrio} and various  surveys (see, for
instance,
Reference \cite{bord} and monographs \cite{mill},\cite{milto}). 

When a vibrating cantilever approaches a sample there is an increase in the
dissipative force it experiences. Because of its great practical importance,
the problem of noncontact friction 
has become a topic of increasing interest (see, for instance, References
\cite{dorprl}-\cite{volper}). The utilization of micromechanical cantilevers
could conceivably be extended considerably
 if the physical mechanisms responsible for the damping were understood. Such an
understanding might lead to applications in which both the conservative
(elastic or electrostatic) forces
and the dissipative (velocity-dependent) forces are probed in order to obtain
useful information about a sample.
 
Attempts to explain the observed friction in terms of the van der Waals
interaction \cite{pendry},\cite{rob}-\cite{bart} have not met with much success
\cite{vol},\cite{perss},\cite{volper}.
On the other hand the effect of electric fields of various origins on the
noncontact friction was shown in the experiments of 
Stipe {\it et al.} \cite{ru} to play an important role in determining the
friction, 
although a microscopic theory still does not exist. It follows from the work of
Stipe {\it et al.} \cite{ru} that any changes in the electric field (due, for
example, to changes of 
tip-surface distance or variations in bias voltage associated with charge
defects generated in dielectric substrates) produce significant modifications
of the noncontact friction.
 
 In what follows we consider the perpendicularly oriented cantilever, which has
been utilized as an ultrasensitive device for the study of forces and
dissipation at small length scales \cite{ru}. 
The perpendicular orientation allows a much closer approach of  the tip to the
sample without jump-to-contact, and also much less cantilever damping than in
the parallel configuration. 
The influence of tip-substrate attraction on cantilever eigenfrequencies and
Joule dissipation will be studied, and the dependence of these quantities on
the tip geometry will be described 
both analytically and numerically. 

In Section II we develop expressions for the electric field , induced
currents, and Joule heating in the case of a cylindrical tip. We consider both
infinitely thick and finite-thickness
substrates, and find that Joule heating is increased in the case of a thin
substrate. Section III extends this
analysis to the more realistic case of spherical or ellipsoidal tips, and in
Section IV we consider the Casimir force,
which can be the dominant force at small tip-sample separations, for both
cylindrical and ellipsoidal tips. In
Section V we consider the force arising from spatial fluctuations of
electrostatic charge, and compare this force to
the Casimir force for typical parameters of interest. Section VI is concerned
with the effects of these attractive forces on the
vibrational frequency and effective mass of the cantilever, and numerical
results for these effects are given for typical values
of various parameters. Our conclusions are summarized in Section VII.

\section{Electric field and induced currents for a cylindrical tip geometry}

\begin{figure}[t]
\centerline{\psfig{file=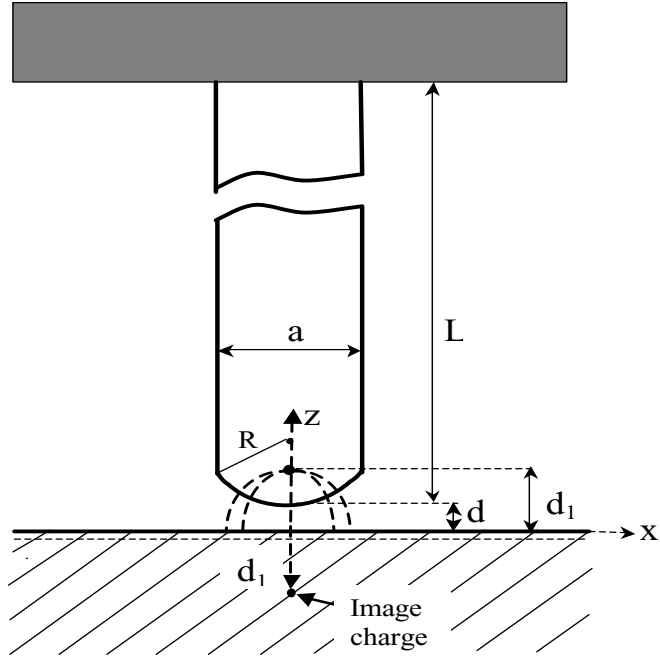,width=9cm,height=10cm,clip=}}
\vspace{4mm}
\caption{Cantilever-sample geometry for the model in which the cantilever ``tip"
is taken to be part of a cylinder of radius $R$.}
\label{fig:1}
\end{figure}

In this section we consider first a model in which the cantilever tip is assumed
to be a metallic cylinder with its axis parallel to a flat sample surface. The
tip displacements are assumed to be 
nearly parallel to the surface, which will be the case when the oscillation
amplitudes are sufficiently small. Figure \ref{fig:1} shows the model under
consideration. The cantilever width $b$
is taken to be much larger than the thickness $a$ ($b\gg a$). Figure \ref{fig:1}
shows a section in the plane of the cantilever 
motion. It is straightforward to obtain the static electric field distribution
in the practically important case of small distances $d$ such that  the
electrostatic field of the entire
 cylinder is effectively the same as that due only to its bottom part. (The
criterion that $d$ must satisfy for this to be the case is given below.) The
problem is then reduced to solving the 
2D Laplace equation with the boundary conditions that the potential has constant
values $\phi_1$ and $\phi_2$ at the metallic surfaces. The electric field
distribution outside the 
conductors is equal to the field due to a charge $q$ and its image $-q$ placed
at a distance $d_1=[(R+d)^2-R^2]^{1/2}$ from the surface. (Details may be found
in Reference
\cite{ele}.) $q$ represents charge per unit length 
along the $y$ direction, and is determined by the potential difference
$\phi_1-\phi_2$ and the capacitance $C$ given by
\be
C^{-1}=2{\rm cosh}^{-1}(1+d/R)=2\ln\left[(d+R+d_1)/R\right]  \ .
\label{eq1}
\ee
Thus \cite{ele}
\be
q=C(\phi_1-\phi_2) \ .
\label{eq2}
\ee
The electric field at a point $\br$ exterior to the tip and sample is given by
\be
{\bf
E}(\br)=2q\left[{\br-\br_1\over|\br-\br_1|^2}-{\br-\br_2\over|\br-\br_2|^2}\right]
\ ,
\label{eq3}
\ee
where $\br_1$ and $\br_2$ denote the positions of  $+q$ and -$q$, respectively.
For the coordinate system shown in Figure \ref{fig:1},
$\br_{1,2}=\pm d_1{\bf e}_z$, where ${\bf e}_z$ is the unit vector along the $z$
axis. The attractive cantilever-surface force can be calculated
straightforwardly
using Eqs. (\ref{eq1})-(\ref{eq3}). We obtain
\be
T=bq^2/d_1 
\label{eq4}
\ee
for the force on the tip. $T$ is proportional to the length $b$ of the
cylindrical tip in our model, and is given by
\be
T={bV^2R^{1/2}\over 2^{7/2}d^{3/2}} \ , \ \ \ \ V=\phi_1-\phi_2
\label{eq5}
\ee
in the practically important case $d\ll R$. 

  Metals are almost inpenetrable by the electric field, the penetration being
restricted to a very thin surface layer (Debye layer) 
of  thickness $R_D$ considerably less than other characteristic lengths of the
system, i.e., $R_D\ll R,d,d_1$. Integrating the Poisson equation over a thin
layer of the sample
(but with  thickness greater than $R_D$), we obtain the surface charge density
\be
\sigma(x)=-{qd_1\over\pi[x^2+d_1^2]} \ ,
\label{eq6}
\ee
where we have taken into account  the boundary conditions
$E_z(z=0^+)=-4qd_1/(x^2+d_1^2)$ and $E_z(z=0^-)=0$. The last condition, of
course, means that the overall field due to 
the charged tip and the surface (Debye) layer is zero in the volume of the metal
substrate. 

A somewhat different picture applies in the case of a {\it moving} charged tip.
To be specific, we consider the case of oscillatory motion of the tip:
$X(t)=X_0\cos\om t$. 
(Note that  an oscillating cantilever is used, for example, for spin detection
\cite{rugar}.) 
The cantilever charge is not changed when its tip moves parallel to the surface,
while  the sample charge varies in time at any fixed point. The varying sample
charge induces 
an electric current, and the charge and current are related by the continuity
equation, $\partial\rho/\partial t+\nabla\cdot{\bf J}=0$, where $\rho$ and
${\bf  J}$ are the charge
and current densities, respectively. Integration of the continuity equation over
a small element of surface layer gives the relation between the normal current
near the Debye layer and 
the time derivative of the surface charge density:
\be
{\pa\over\pa t}\sigma(x,t)=J_z(x,z=0^-,t) \ .
\label{eq7}
\ee
We will assume that the cantilever motion is slow in the sense that $\om\ll
4\pi\mu$, where $\mu$ is the sample conductivity; in 
this case an iterative procedure described by
Boyer \cite{timo} is applicable. The assumption of slow cantilever motion means
that the surface charge follows the tip displacements practically
instantaneously. To lowest order in 
the parameter $\om/(4\pi\mu)$, the surface charge may be approximated by Eq.
(\ref{eq6}) with the coordinate $x$ replaced by $x-X(t)$. This approximation
corresponds to an 
effectively infinite value of volume conductivity and instantaneous recharging.
Then, from  Eq. (\ref{eq7}),
\be
J_z(x,z=0^-,t)=\mu E_z(x,z=0^-,t)=-{1\over\pi}q d_1\om X_0\sin\om t{\pa\over\pa
x}{1\over (x-X)^2+d_1^2} \ .
\label{eq8}
\ee

The additional surface charge $\sigma'$ (up to first order in $\omega/mu$) 
caused by the normal current (\ref{eq8}) can easily be obtained by considering
this charge as a source of the field $E_z(x,z=0^-,t)$. 
Taking into account the fact that the field $E_z$ induced by the additional
surface charge has opposite signs below and above the surface $z=0$, we obtain
the perturbation of  the surface 
charge density as follows:
\be
\sigma'(x,t)={qd_1\om X_0\sin\om t\over 2\pi^2\mu}{\pa\over\pa
x}{1\over(x-X)^2+d_1^2} \ .
\label{eq9}
\ee  
For small tip oscillation amplitudes $(X_0\ll d_1)$ we can ignore $X(t)$ in the
denominator. Thus we obtain the electric field due to the surface charge
$\sigma'$:
\be
{\bf E}'(\br,t)=2\int_{-\infty}^{\infty}
dx'\sigma'(x',t){\pa\over\pa\br}\ln|\br-{\bf e}_xx'| \ .
\label{eq10}
\ee                  
Explicitly, the expression for the electric field that is first-order in
$\omega/\mu$ is
\be
{\bf E}'(\br,t)={q\om X_0\sin\om t\over \pi\mu}{\pa\over\pa d_1}\left({-{\bf
e}_x(|z|+d_1)\pm {\bf e}_zx\over x^2+(|z|+d_1)^2}\right) \ ,
\label{eq11}
\ee
where the $+$ and - signs correspond to $z>0$ and $z<0$, respectively. It
follows from Eq. (\ref{eq11}) that the characteristic length of the electric
field variation is $d_1$. Hence our simplifying assumption that the field formed
by the full cylinder can be represented by only its bottom part
(as in Figure \ref{fig:1}) is valid in the case $d_1\ll R$.

Equation (\ref{eq11}) can be used to obtain the cantilever damping due to Joule
dissipation. The Joule power $W_j$ is given by the integral 
\be
W_j={b\mu}\int_{-\infty}^{\infty} dx\int_{-\infty}^{0} dz\left[{\bf
E}'(\br,t)\right]^2 \ ,
\label{eq12}
\ee
which, when used with Eq. (\ref{eq11}), becomes
\be
W_j={b(qX_0\om)^2\over 8\pi\mu d_1^2}\approx {b(VX_0\om)^2\over 2^7\pi\mu d^2} \
.
\label{eq13}
\ee 
Note that there is no dependence of $W_j$ on the tip radius $R$. This results
from the effects of two competitive factors determining the Joule loss. The
first is the magnitude of the electric 
field in the bulk of the substrate, which increases with a decrease in the
characteristic inhomogeneity size $d_1\approx (2dR)^{1/2}$ as $R/d_1^3$. The
second is the characteristic volume 
$\sim d_1^2$ contributing to the integral in Eq. (\ref{eq12}). Thus the two
effects together vary as $(R/d_1^3)^2d_1^2\propto 1/d^2$, and have no net
dependence on $R$.

 The damping of the cantilever oscillations can be described alternatively as
follows. It can be seen from Eq. (\ref{eq11}) that there exists not only a
$z$-component, but also an
$x$-component of the field ${\bf E}'$,  i.e., a component parallel to the
direction of tip motion that reduces the cantilever velocity. The loss $W$ of
the kinetic energy per unit time is 
proportional to a friction force and the velocity $\pa X(t)/\pa t$:
\be
W=-bq\overline{E_x'(x=0,z=d_1,t)\pa X(t)/\pa t} \ ,
\label{eq14}
\ee
where the overline denotes an average over a period of the oscillations. The
explicit form of $W$ can be obtained using Eq. (\ref{eq11}), and it is easily
verified that $W$
obtained in this way is exactly the same as that given by Eq. (\ref{eq13}),
i.e., $W=W_j$.

\begin{figure}[t]
\vspace{-3cm}
\centerline{\psfig{file=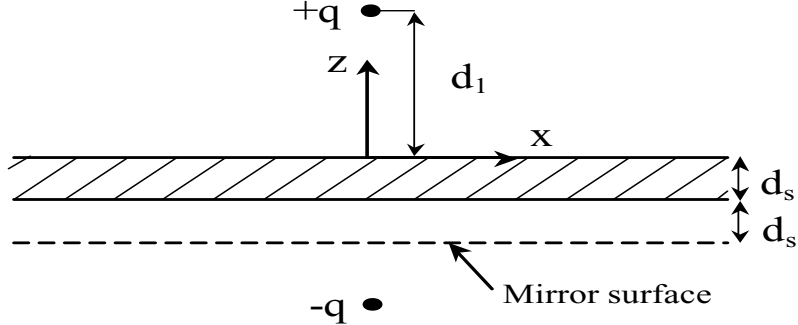,width=11cm,height=11cm,clip=}}
\vspace{-3.4cm}
\caption{Cantilever-sample geometry for the case 
of finite sample thickness.}
\label{fig:2}
\end{figure}

In the case of finite sample thickness $d_s$ we assume that the Debye layer
thickness $R_D$ is negligible. (For good metals, this 
thickness is of order 1 $\AA$.) 
For a finite sample thickness we must account for the additional boundary
condition at $z=-d_s $ (see Fig. \ref{fig:2}). To derive this boundary
condition, we
use the fact that when $d_s\gg R_D$ the static field does not penetrate to the
depth $d_s$. An electric charge can be induced at the second boundary of the
sample only by virtue of the field 
${\bf E}'\propto\om$. Therefore the term $\pa\rho/\pa t$ in the continuity
equation can be neglected, as it is of order $\om^2$, so that the boundary
condition at $z=-d_s$
 for zero surface charge is
\be
j_z(x,z=-d_s)=\mu E_z(x,z=-d_s)=0 \ .
\label{eq15}
\ee  
Thus the solution of the reduced continuity equation $\nabla\cdot{\mu\bf E}'=0$
with boundary conditions (\ref{eq8}) and (\ref{eq15}) determines the field
${\bf E}'$
in the bulk of the sample. Further analysis becomes more convenient if we
consider the field ${\bf E}'$ to be induced by charges on the surface $z=0$ and
on the ``image plane"
at $z=-2d_s$. Setting the charge density  at the image plane equal to that at
the $z=0$ plane, i.e., $\sigma'(x,z=0)=\sigma'(x,z=-2d_s)$, ensures that the
boundary condition 
$j_z(x, z=-d_s)=0$ is satisfied. The electric field ${\bf E}'$ can be expressed
in terms of $\sigma'$ as
\be
{\bf E}'(\br,t)=2\int_{-\infty}^{\infty}
dx'\sigma'(x',t){\pa\over\pa\br}[\ln|\br-{\bf e}_xx'|+\ln|\br-{\bf
e}_xx'+2d_s{\bf e}_z |] \ .
\label{eq16}
\ee

To obtain an expression for $\sigma'$ we use Eqs. (\ref{eq8}) and (\ref{eq16}):
\bea
-{1\over\pi\mu}&&{qd_1\om X_0\sin\om t}{\pa\over\pa x}\left({1\over
x^2+d_1^2}\right)=\nonumber\\
&& 2\int dx'\sigma'(x',t)  
\left[{2d_s-\Delta\over(2d_s-\Delta)^2+(x-x')^2}-
{\Delta\over\Delta^2+(x-x')^2}\right]_{\Delta\rightarrow 0} \ .
\label{eq17}
\eea
The solution of this equation for $\sigma'(x,t)$, for the practically important
case of a thin sample ($d_s\ll d_1$), is
\be
\sigma'(x,t)=-q\om X_0\sin\om t{x\over 4\pi^2\mu d_s(x^2+d_1^2)} \ .
\label{eq18}
\ee
The field component responsible for friction, $E_x'(x=0,z=d_1,t)$, then follows
from Eqs. (\ref{eq16}) and (\ref{eq18}):
\be
E_x'(x=0,z=d_1,t)={q\om X_0\sin\om t \over 2\pi\mu d_1d_s} \ .
\label{eq19}
\ee
This is seen to be greater by a factor of $2d_1/d_s$ than the corresponding
field for the infinitely thick sample. Therefore the Joule
loss $W_j^{FS}$ is increased by the same factor:
\be
W_j^{FS}={b(VX_0\om)^2R^{1/2}\over 2^{11/2}\pi\mu d^{3/2}d_s} \ .
\label{eq20}
\ee
The increase of Joule dissipation in a thin substrate is due to the decrease of
the characteristic volume where recharging occurs. 
In this case the effective resistivity of the substrate is increased, giving
rise to additional ohmic dissipation. 

\section{Spherical and ellipsoidal tips}

\begin{figure}[t]
\centerline{\psfig{file=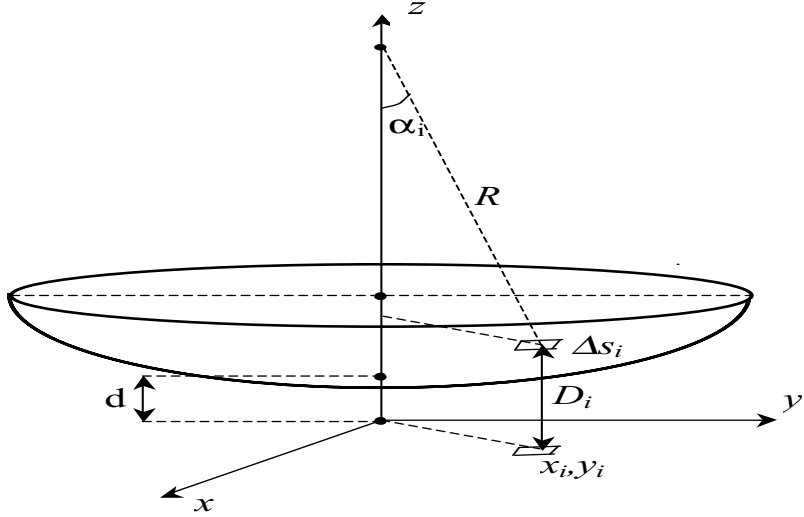,width=11cm,height=9cm,clip=}}
\vspace{4mm}
\caption{In the ``proximity approximation" the force is obtained by adding the
pairwise forces $\Delta f_i$ between
opposing surface elements separated by distance $D_i$.}
\label{fig:3}
\end{figure}

We now consider the case of a spherical or ellipsoidal tip.  As before we assume
a tip-plane separation much smaller than the curvature radius $R$, allowing
us to make a ``proximity approximation" as follows. For two {\it parallel}
surfaces (the limit $R\rightarrow\infty$) the electric field has only a $z$
component, and the attractive force 
between any pair of opposing elements is
\be
\Delta f_i=\Delta s_if(d)=\Delta s_i{V^2\over 8\pi d^2} \ .
\label{eq21}
\ee
For large (but finite) $R$ we may assume that Eq. (\ref{eq21}) holds, but with
$d$ replaced by the distance $D_i$ between opposing elements $\Delta s_i$ (see
Figure \ref{fig:3}). It follows from elementary geometrical considerations that
for small angles $\alpha_i$ the distance $D_i$ is given by
\be
D_i\approx d+(R/2)\alpha_i^2 \ ,
\label{eq22}
\ee
and the total force acting on the tip is  
\be
T=\sum_i\Delta f_i\approx{V^2R^2\over 4}\int{d\alpha\alpha\over D^2(\alpha)} \ .
\label{eq23}
\ee
Since small angles $\alpha$ $\sim(d/R)^{1/2}$ make the dominant contribution to
the integral,  the integration range can be taken from 0 to $\infty$.
Thus we have
\be
T={RV^2\over 4d} \ .
\label{eq24}
\ee
It is worth mentioning that the application of this ``proximity approximation"
approach to the case of the cylindrical tip gives exactly the force of
attraction (\ref{eq5}).

 Let us consider now an ellipsoidal  tip surface having two different curvatures
in perpendicular directions. In the vicinity of a 
sample the equation describing this surface involves a quadratic dependence on
$x$ and $y$:
\be
z(x,y)={x^2\over 2R_x}+{y^2\over 2R_y} \ ,
\label{eq25}
\ee
where $R_x$ and $R_y$ are the radii of curvature in the two directions. Our
previous considerations may be repeated with the modification that the
tip-surface distance
$D_i$ is now taken to be
\be
D_i=d+z(x_i,y_i) \ .
\label{eq26}
\ee
 For the purpose of integration it is useful to introduce the change of
variables 
\be
x=(R_x/R_y)^{1/2}x' \ , \ \ \ \ y=(R_y/R_x)^{1/2}y' \ ,
\label{eq27}
\ee
in terms of which the problem is effectively reduced to the spherical case, and
the force for the ellipsoidal tip is given by Eq. (\ref{eq24}) with
$R=(R_xR_y)^{1/2}$.
  
The calculations of Joule loss in the case of a spherical tip proceed as
outlined in the previous section. The surface charge density of the sample for
the moving cantilever now depends on both $x$ and $y$:
\be
\sigma(x,y,t)={VR/2\pi\over 2dR+[x-X(t)]^2+y^2} \ .
\label{eq28}
\ee                               
The  additional surface charge $\sigma'$ that generates dissipative currents in
the sample is
 \be
\sigma'(x,y)=-{1\over 2\pi\mu}{\pa\over\pa t}\sigma(x,y,t) \ ,
\label{eq29}
\ee                              
and the electric field responsible for dissipation is determined from $\sigma'$
by
\be
{\bf E}'(\br,t)=-\nabla\int
d{\br}_{\perp}{\sigma'({\br}_{\perp})\over|\br-\br_{\perp}| } \ ,
\label{eq29}
\ee
where $\br=(x,y,z)$ and ${\br}_{\perp}=(x,y)$. The expression for the Joule loss
is 
\be
W_j={(VX_0\om)^2R^{1/2}\over 32\sqrt{2}\pi^3\mu d^{3/2}}I \ ,
\label{eq30}
\ee where the (dimensionless) integral $I$ is given by 
\be
I=\int\int d\br_{\perp}d\br'_{\perp}|\br_{\perp}-\br'_{\perp}|^{-1}{\pa\over\pa x}
(1+r^{2}_{\perp})^{-1}{\pa\over\pa x'}(1+r^{'2}_{\perp})^{-1} \approx 4.8 \ .
\label{eq31}
\ee
The $d^{-3/2}$ dependence of $W_j$ on $d$ given by Eq. (\ref{eq30}) is somewhat
weaker than in the cylindrical case, and is comparable to the dependence on
$d$ of the frictional force observed by Stipe, {\it et al.} \cite{ru}. However,
the numerical value given by 
Eq. (\ref{eq30}) is too small to explain the experimental results. This question
will be taken up in Section 6.

\section{Casimir force}

The ``proximity approximation" employed in the previous section can be applied
to the case of the Casimir attraction. We consider the case of small
separations $d$
such that $d\ll c/\om_p,\hbar c/k_BT$, where $\om_p$ is the plasma frequency;
$\om_p\approx 10^{16}$ s$^{-1}$ for good metals, so that we require $d\ll 30$
nm. For these
small separations we can use the Lifshitz formula \cite{lif} for the force per
unit area:
\be
f(d)={\hbar\over 8\pi^2d^3}\int d\xi{[\eps(i\xi)-1]^2\over[\eps(i\xi)+1]^2} \ ,
\label{eq32}
\ee
where $\eps$ is the dielectric permittivity, assumed to be the same for both
plates. The explicit form of $\eps$ in the Drude approximation is 
\be
\eps(i\xi)=1+{\om_p^2\over\xi(\xi+\nu)} \ ,
\label{eq33}
\ee
 where $\nu$ is the electron collision frequency ($\approx 10^{13}$ s$^{-1}$ at
$T=300 $K), which is much smaller than $\om _p$, and can be neglected when
integrating Eq. (\ref{eq32}). It follows from Eqs. (\ref{eq32}) and
(\ref{eq33}) that
\be
f(d)={\hbar\om_p\over 32\sqrt{2}\pi d^3} \ .
\label{eq34}
\ee                                         
Using (\ref{eq34}), (\ref{eq21}), and (\ref{eq23}), we obtain
\be
T(d)={R\hbar\om_p\over 32\sqrt{2}d^2} 
\label{eq35}
\ee
for a spherical tip of radius $R$. In the case of a cylindrical tip we have
\be
T(d)={3bR^{1/2}\hbar\om_p\over 2^8d^{5/2}} \  .
\label{eq36}
\ee 
This differs from the force obtained for the case of a cylinder in Reference
\cite{dorprb} ($T(d)\propto d^{-(n-1)}, \ n=3$ or 4). The 
latter result was obtained using the unjustified assumption that the Casimir
force satisfies Laplace's equation in the gap between the metals. 
It is seen from Eqs. (\ref{eq35}) and (\ref{eq36}) that the dependence of the
Casimir force on $d$ is stronger 
than in the electrostatic case. At small values of $d$ the Casimir force can be
dominant.

\section{Attraction due to spatial fluctuations of electrostatic charge}

In this section we consider a substrate with a stationary, inhomogeneous
distribution of the electric charge, which 
can occur in good dielectrics. Such a situation was approximated experimentally
\cite{ru} by employing a fused silica sample 
irradiated with $\gamma$ rays. In the course of irradiation, positively charged
centers (Si dangling bonds) are generated. 
Randomly distributed positive charges are embedded in a negative background, and
on average the sample is electrically neutral. 
We model the sample as consisting of microscopically small volume elements
$\Delta V_i$. Each element is chosen sufficiently 
small that not more than one charge center is present in it. The electric charge
$q_i$ of each element can be described by 
\be
q_i=n_ie+(1-n_i)e_b \ ,
\label{eq37}
\ee
where $n_i=1$ or 0 for an occupied or unoccupied charge center element,
respectively, and $e_b$ is the background charge in the
 volume $\Delta V_i$. It follows from Eq. (\ref{eq37}) that the condition of
charge neutrality, $\la q_i\ra=0$, is satisfied when 
\be
e_b=-{e\la n_i\ra\over 1-\la n_i\ra}\approx -en \ ,
\label{eq38}
\ee
where $n\equiv \la n_i\ra$,  $n\ll 1$. In the following we will consider the
fluctuations of charges in different volume elements $i, j$ 
to be  statistically independent, so that  $\la\delta q_i\delta g_j\ra=0$ for
$i\neq j$. The mean square of charge fluctuations within
a given volume element is given by
\be
\la\delta q_i\delta q_i\ra=\la n_i^2\ra e^2+2ee_b\la n_i(1-n_i)\ra
+\la(1-n_i)^2\ra e_b^2 \ .
\label{eq39}
\ee
It follows from the definition of $n_i$ that $n_i^2=n_i$, $(1-n_i)^2=1-n_i$, and
$n_i(1-n_i)=0$. Then using the assumption that $n\ll 1$, we obtain from Eq.
(\ref{eq39}) the mean square
\be
\la\delta q_i\delta q_i\ra\approx ne^2 \ ,
\label{eq40}
\ee
which shows that a random charge distribution is consistent with Poisson
statistics.
Each individual charge $q_i$ has its own image $-q_i$ in the bulk of the
cantilever tip. Then, in the absence of cross terms, 
an average tip-sample attractive force is determined by adding the contributions
from all the charge-image pairs. 
Thus the electrostatic attraction $\Delta f_i$ between opposing elements $\Delta
V_i$ is given by
\be
\Delta f_i={ne^2\over 4[D(x_i,y_i)-z_i]^2}  \ ,
\label{eq41}
\ee
where the coordinates $x_i$, $y_i$, $z_i$ give the position of the $i$th volume
element in the substrate, and the total force is obtained 
by summing over all the elements:
\be
T={ne^2\over 4}\sum_i{1\over [D(x_i,y_i)-z_i]^2} \ .
\label{eq42}
\ee
After replacing the sum by an integral ($n\sum\rightarrow c\int d^3r$, where $c$
is the number of charge centers per unit volume), 
we obtain, after integration over $z$,
\be
T={ce^2\over 4}\int\int {dxdy\over D(x,y)}={\pi ce^2R^2\over 2d}\int
{d\alpha\alpha\over 1+{\alpha^2R\over 2d}} \ .
\label{eq43}
\ee
The integral on the right-hand side is logarithmically divergent. Hence we
should confine the integration to values of $\alpha$
up to $\alpha_{max}\approx R_{tip}/2R$, where $R_{tip}$ is the characteristic
size of the bottom part of the tip. The value of 
the integral is not very sensitive to $R_{tip}$, so that we may set $R=R_{tip}$.
Then the force of attraction is given by
\be
T={1\over 2}\pi ce^2R\ln\left[1+{R\over 8d}\right] \ .
\label{eq44}
\ee

  To illustrate the importance of this force, we put in Eq. (\ref{eq44}) the
typical values $c=10^{18}$ cm$^{-3}$ (as in \cite{ru,rugar}, see also
references therein) for the
defect concentration, and $R\approx 1$ $\mu$m, $d=20$ nm. In this case the ratio
of the force (\ref{eq44}) to the Casimir force (\ref{eq35}) is $10$. 
This ratio decreases with decreasing  $d$. 

\begin{figure}[t]
\centerline{\psfig{file=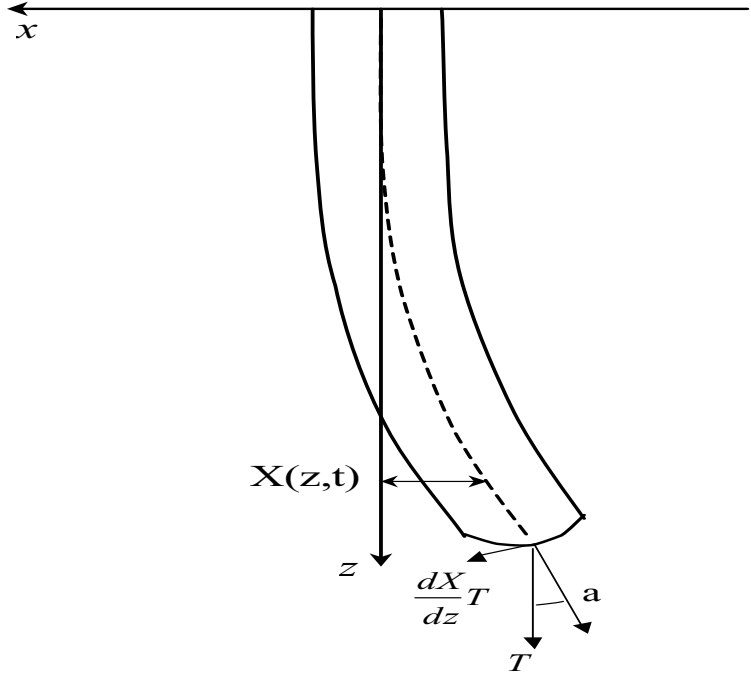,width=11cm,height=9cm,clip=}}
\vspace{4mm}
\caption{Transverse displacement $X(z,t)$ of the cantilever centerline.
Restoring force $X^\prime T$ acts on the tip when $\alpha \neq 0$.}
\label{fig:4}
\end{figure}

Our considerations in this section have ignored the screening of the electric
field in the dielectric substrate. This can be justified in the case of very
small tip-sample separations (substantially smaller than screening length), as
only defects 
in the surface layer of  thickness $d$ contribute to the integral in Eq.
(\ref{eq43}). When the screening is important, the effective 
electric field inducing image charge in the metal is decreased by the factor of
$(\eps +1)/2$ (see Reference \cite{ele}). This
factor is equal to 2.5 in the case of silica. In addition, the tip image charge
induces another image charge in the dielectric, and so 
forth. Nevertheless, for rough estimates it is sufficient to consider only the
first pair of charges, equal to $\pm {2e/(\eps +1)}$. 

\section{Influence of attractive force on cantilever eigenfrequencies}

When the cantilever geometry satisfies the conditions $L\gg b\gg a$ we can
employ the equation of elasticity \cite{ela}
 to describe the dynamics. In the case of an isotropic material, the transverse
displacement $X(z,t)$ of the beam centerline (along the $x$ axis, see Figure
\ref{fig:4}) satisfies the differential equation 
\be
\rho_cS{\pa^2\over\pa t^2}X(z,t)=-EI_i{\pa^4\over\pa z^4}X(z,t) \ ,
\label{eq45}
\ee  
where $\rho_c$ is the density of the cantilever material, $S=ab$ is the
cross-section area, $E$ is Young's modulus, and
$I_i =a^3b/12$ is the bending moment of inertia. The clamped end, at $z=0$,
imposes the boundary conditions 
$X(z=0)=X'(z=0)=0$. The free end at $z=L$ imposes the boundary conditions
$M_y\equiv EI_iX''=0$ and $F_x\equiv -EI_iX'''=0$,
where $M_y$ and $F_x$ are the moment of the elastic force and the elastic force
in the $y$ and $x$ directions, respectively. 
The fundamental eigenfunction is given by $X(z,t)=X_0(z)\cos(\om t)$, where
\be
X_0(z)=A[(\cos\kappa L+\cosh\kappa L)(\cos\kappa z-\cosh\kappa z)+(\sin\kappa L-\sinh\kappa L)
(\sin\kappa z-\sinh\kappa z)] \ .
\label{eq46}
\ee
Here $A$ is a constant, $\kappa^4=\om^2\rho_cS/EI_i$, and
\be
\om=(3.516/L^2)(EI_i/\rho_cS)^{1/2} \ .
\label{eq47}
\ee
 To estimate $\om$ we put $a=250$ nm, $b=7\times 10^3$ nm, $L=250$ $\mu$m, and
use the material constants of Si:
$\rho_c=2.33$ g cm$^{-3}$, $E=9\times 10^{11}$ g cm$^{-1}$ s$^{-2}$. Then we
have $\om/2\pi\approx 4\times 10^3$ s$^{-1}$,
which is approximately the same as the value quoted in Reference \cite{ru}.

In the presence of a stretching force $T$, Eq. (\ref{eq45}) is modified to
\cite{ela}
\be
\rho_cS{\pa^2\over\pa t^2}X(z,t)=-EI_i{\pa^4\over\pa z^4}X(z,t) +T{\pa^2\over\pa
z^2}X(z,t) \ .
\label{eq48}
\ee                   
Furthermore the boundary condition at $z=L$ ($X'''=0$) should be replaced by the
more general condition 
$-EI_iX'''+TX'=0$. The term $TX'$ describes the effect of the restoring force
when there is a nonvanishing angle between $T$
  and the centerline direction at $z=L$ (see Figure \ref{fig:4}). This value of
the restoring force is valid in the case of small angles 
$\alpha$, when $\sin\alpha\approx\tan\alpha=-X'(L)$.

 The eigenfunctions of Eq. (\ref{eq48}) are determined by
\bea
X_0(z)&=&A[(\Lambda^2\cos\lambda+\lambda^2\cosh\Lambda)(\cos\lambda
z-\cosh\Lambda z) \nonumber \\ 
&+& (\Lambda\sin\lambda-\lambda\sinh\Lambda)(\Lambda\sin\lambda
z-\lambda\sinh\Lambda z)] , 
\label{eq49}
\eea
where 
\be
\Lambda=L[(\tau/2)^2+(\kappa^4+\tau^2/4)^{1/2}]^{1/2} \ , \ \ \ 
\lambda=L[(-\tau/2)^2+(\kappa^4+\tau^2/4)^{1/2}]^{1/2} \ , \ \ \ 
\tau=T/EI_i \ .
\label{eq49A}
\ee
For simplicity the coordinate $z$ in Eq. (\ref{eq49}) is given in units of $L$. 
The value of $\kappa$ in Eq. (\ref{eq49A}) obeys the following eigenvalue
equation:
\be
2\kappa^4+(\tau^2+2\kappa^4)\cos \lambda\cosh \Lambda+\kappa^2\tau\sin\lambda
\sinh\Lambda=0 \ .
\label{eq50}
\ee 
The minimum value of $\kappa$ ($\kappa_{min}$) that obeys Eq. (\ref{eq50})
determines the fundamental eigenfrequency and the explicit form of the
fundamental eigenfunction (\ref{eq49}).

\begin{figure}[t]
\vspace{-1.5cm}
\centerline{\psfig{file=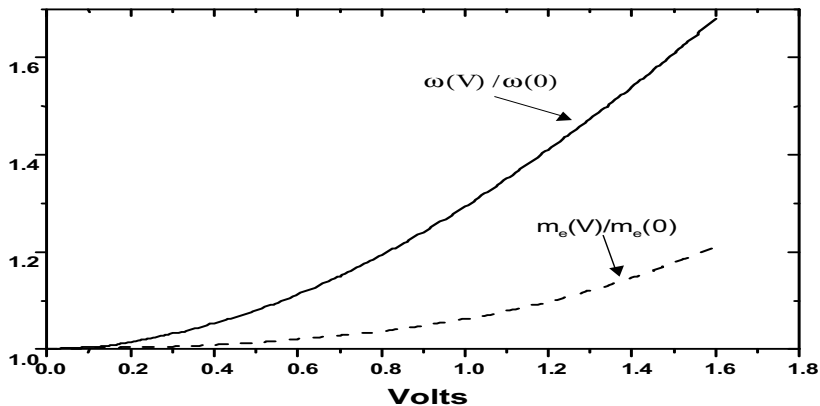,width=11cm,height=9cm,clip=}}
\vspace{-1.5cm}
\caption{Numerical results for the cantilever frequency $\omega$ and effective mass
$m_{eff}$  as a function of the bias voltage $V$ in the case of a cylindrical
tip (see text).}
\label{fig:5}
\end{figure}

In the general case the fundamental frequency is a function of  the tip-sample
separation, bias voltage $V$, tip geometry, and 
concentration of charge centers via the attraction force $T(V)$. In the case of
attraction due only to an external bias voltage,
there is a quadratic dependence of $T$ on $V$. The proportionality coefficient
depends on the geometry of the tip-sample system. 
To illustrate the $\om(V)$ dependence we use the parameters given at the
beginning of this section. The results of numerical
 calculations are shown in Figure \ref{fig:5} for a cylindrical geometry and for
the same parameters assumed as at the beginning of this section. It is seen
that the bias voltage can have a considerable effect on the frequency shift.

 In the limit of very large $V$ (when $TL^2/(EI_i)\gg 1$), the first term on the
right-hand side of Eq. (\ref{eq48}) can be omitted, and Eq. (\ref{eq48}) is
transformed into the equation for the vibration
of a string. In this case, a linear dependence of $\om$ on $V$ ($\om\propto
V/L$) should occur. 

 Along with the resonant frequency, the cantilever effective mass also depends
on $V$. The effective mass $m_{eff}$ is the coefficient entering the oscillator
equation
\be
m_{eff}{\pa^2X(t)\over\pa t^2}+\Gamma{\pa X(t)\over\pa t}+kX(t)=0
\label{eq51}
\ee
as a parameter that determines the inertial force. ($\Gamma$ is a friction
coefficient, $k$ a spring constant, and $X(t)$, as previously, is the
coordinate of the tip.) The mass $m_{eff}$ is 
related to the coefficients in Eq. (\ref{eq48}). This relation can be obtained
from the requirement that the kinetic energy of the flexible beam equals that
of the oscillator, the two being
one and the same physical quantity. This condition defines $m_{eff}$ as
\be
m_{eff}=\rho_cS\int_0^LdzX_0^2(z)/X_0^2(L) \ ,
\label{eq52}
\ee             
where $X_0(z)$ is given by Eq. (\ref{eq49}).

Thus $m_{eff}$  depends on the variation of $X$ with $z$, which in turn depends
on the stretching force $T$ and the other parameters in Eq. (\ref{eq48}). In
the simplest case $T=0$, 
we have $m_{eff}=\rho_cLS/4$, i.e., a quarter of the beam's actual mass. In
general the possible variation of $m_{eff}$ should be taken into account when
the friction coefficient $\Gamma$ is obtained experimentally from the  relation
$\Gamma=m_{eff}\om/Q$, where $Q$ is the quality factor. In other words, not
only the quality factor and the eigenfrequency, but also the value of $m_{eff}$
is required to obtain $\Gamma$. The dependence $m_{eff}(V)$ is shown in Figure
5. It is seen to be similar to that of $\om(V)$ at small values of $V$.

\begin{figure}[t]
\vspace{-1.5cm}
\centerline{\psfig{file=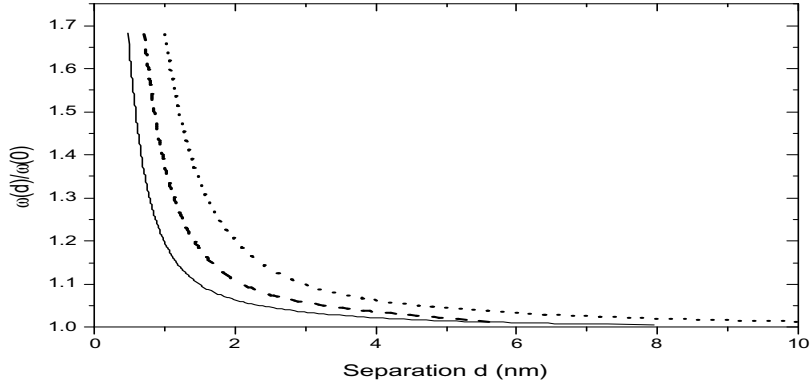,width=11cm,height=9cm,clip=}}
\vspace{-1.5cm}
\caption{Effect of Casimir force on the frequency $\om$. Solid line: $R=0.5$ $\mu
$m; 
dashed line: $R=1$ $\mu $m; dotted line: $R=2$ $\mu $m}
\label{fig:6}
\end{figure}

Figure \ref{fig:6} shows  the effect of the Casimir force on the fundamental
frequency in the case of a spherical tip. The three curves illustrate the
tendency of $\om(d,R)$ to increase with the attractive force,
which varies with $d$ and $R$ as $T\sim R/d^2$  (see Eq. (\ref{eq35})). The
frequency shift observed in Reference \cite{ru} at $d=2$ nm for a gold sample
corresponds to the curve $R=0.5$ $\mu$m in Figure 6. When we use the
experimental value 1 $\mu$m of the  radius of curvature in the $y$ direction, and
set the radius of curvature in the $x$ direction equal to the cantilever
thickness $a=0.25$$\mu$m, we obtain $R_{exp}=(R_xR_y)^{1/2}=0.5$ $\mu$m., i.e.,
$R_{exp}=R$. Hence the frequency shift
observed by Stipe {\it et al.} \cite{ru} (see Figure 2c of Reference \cite{ru})
may be attributable to the Casimir effect.

 It is interesting to compare the experimental and theoretical values of 
$\Gamma$. The Joule loss $W_j$ [see Eqs. (\ref{eq13}), (\ref{eq20}), and
(\ref{eq30})] determines the friction 
coefficient: It follows from energy conservation and Eq. (\ref{eq51}) that
\be
\Gamma=\frac{2W_j}{\om^2X_0^2}.
\label{eq53}
\ee
In the case of the cylindrical tip we have, from Eqs. (\ref{eq14}) and
(\ref{eq53}),
\be
\Gamma=\frac{bV^2}{2^6\pi\mu d^2} \ .
\label{eq54}
\ee
 Putting, as previously, $b=7\times 10^{-4}$ cm, a conductivity $\mu=4\times
10^{17}$ s$^{-1}$ for gold at 300 K, $d=20$ nm, and $V=1$ Volt,
we have $\Gamma=2.4\times 10^{-20}$ kg/s, which is much smaller than the
experimental value of $3\times 10^{-12}$ kg/s.
For the more realistic case of a spherical tip with $R=1$ nm, our estimate for
$\Gamma$ using Eq. (\ref{eq30}) is two orders of
magnitude smaller than that given by (\ref{eq54}).

\section{Conclusions}

The analyses presented in this paper indicate that both electrostatic forces
and Casimir forces can have a strong effect on the cantilever vibrations. The
cantilever eigenfrequencies depend on 
the attractive force $T$, which is very sensitive to the tip geometry for small
values of the tip-sample separations $d$.  The effect of various forces on the
cantilever
eigenfrequencies appears to us to be an important consideration for the
practical utilization of cantilever-based devices. It has served as a
motivation in this paper to study the tip-sample 
interaction in some detail.

 We have shown that, for small separations, $T$ depends on the shape of the
cantilever tip. The dependence of $T$ on the separation $d$ is different for
cylindrical
 and spherical tips. Thus, in the case of the electrostatic force due to a bias
voltage, the attractive force varies as $d^{-3/2}$ and $d^{-1}$ for cylindrical
and spherical tips, respectively.
In the case of the Casimir force the dependence of $T$ on $d$ was found to be
$d^{-5/2}$ and $d^{-2}$, respectively, for the cylindrical and spherical tips.
We have shown that the Casimir force is responsible for the frequency shift, which
is of the same order as that obtained experimentally in \cite{ru} ($\approx
4.5$ \%) for a gold sample at small separation $d=2$ $nm$. 

 The attractive force depends furthermore on the radii of curvature, and our
analysis allows for the possibility that the radii may be different in
different directions. 
 In principle, any variations of tip geometry (for instance, due to adsorption
of new molecules or blunting after contact with a sample) may be detected by
measurements of the frequency shift, 
and our theory, which connects the variations of the attractive force with the
frequency shift, could be helpful in estimating the character and scale of
these variations. 

  We have calculated the attractive force between a system of randomly
distributed positive charges, embedded in a negatively charged background, and
a metal tip. In the case of an
electrically neutral system, only the attraction between each charge and its
image contributes to the total force. This is a spatially fluctuating
interaction because the
overall force is linear (not quadratic) in the concentration of elementary
charges. Our estimates, based on 
the assumption of an uncorrelated distribution of charge centers in the bulk of
the substrate, indicate that 
the fluctuating force exceeds the Casimir force for parameters appropriate
to the experiments reported in Reference \cite{ru}, which employed irradiated
silica as a substrate. 

 We have derived analytical expressions for Joule losses in a metal
substrate. The Joule mechanism does not explain the cantilever damping measured in
\cite{ru}, but our analysis may nevertheless be useful for other systems with
high-resistivity substrates. 

\section*{Acknowledgments}
This work  was supported by the Department of Energy under the contract
W-7405-ENG-36 and DOE Office of Basic Energy Sciences,  and by the DARPA
Program MOSAIC.

\end{document}